\documentclass[fleqn,twoside]{article}
\usepackage{espcrc2}
\usepackage{epsfig}
\usepackage[figuresright]{rotating}

\newcommand{\AmS}{{\protect\the\textfont2
  A\kern-.1667em\lower.5ex\hbox{M}\kern-.125emS}}
\newcommand{\ltsimeq}{\raisebox{-0.6ex}{$\,\stackrel 
        {\raisebox{-.2ex}{$\textstyle <$}}{\sim}\,$}} 
\newcommand{\gtsimeq}{\raisebox{-0.6ex}{$\,\stackrel 
        {\raisebox{-.2ex}{$\textstyle >$}}{\sim}\,$}} 
\title{Cosmology with the SKA}

\author{
C.A. Blake\address{School of Physics, University of New South
Wales, Sydney, NSW 2052, Australia},
F.B. Abdalla\address[Ox]{Astrophysics, Department of Physics,
University of Oxford, Keble Road, Oxford, OX1 3RH, UK},
S.L. Bridle\address{Department of Physics and Astronomy, University
College London, London, WC1E 6BT, UK},
S. Rawlings\addressmark[Ox]
}
       
\begin{document}

\begin{abstract}
We argue that the Square Kilometre Array has the potential to make
both redshift (HI) surveys and radio continuum surveys that will {\it
revolutionize} cosmological studies, provided that it has sufficient
instantaneous field-of-view that these surveys can cover a hemisphere
($f_{\rm sky} \sim 0.5$) in a timescale $\sim 1 ~ \rm yr$.  Adopting
this assumption, we focus on two key experiments which will yield
fundamental new measurements in cosmology, characterizing the
properties of the mysterious dark energy which dominates the dynamics
of today's Universe.  Experiment I will map out $\sim 10^9 (f_{\rm
sky}/0.5)$ HI galaxies to redshift $z \approx 1.5$, providing the
premier measurement of the clustering power spectrum of galaxies:
accurately delineating the acoustic oscillations and the `turnover'.
Experiment II will quantify the cosmic shear distortion of $\sim
10^{10} (f_{\rm sky}/0.5)$ radio continuum sources, determining a
precise power spectrum of the dark matter, and its growth as a
function of cosmic epoch.  We contrast the performance of the SKA in
precision cosmology with that of other facilities which will, probably
or possibly, be available on a similar timescale.  We conclude that
data from the SKA will yield transformational science as the direct
result of four key features: (i) the immense cosmic volumes probed,
exceeding future optical redshift surveys by more than an order of
magnitude; (ii) well-controlled systematic effects such as the narrow
`$k$-space window function' for Experiment I and the accurately-known
`point-spread function' (synthesized beam) for Experiment II; (iii)
the ability to measure with high precision large-scale modes in the
clustering power spectra, for which nuisance effects such as
non-linear structure growth, peculiar velocities and `galaxy bias' are
minimised; and (iv) different degeneracies between key parameters to
those which are inherent in the Cosmic Microwave Background.
\end{abstract}

\maketitle

\section{Background}
\label{secback}

Over the last few years we have entered an `era of precision
cosmology' in which rough estimates of many of the key cosmological
parameters have been replaced by what can reasonably be described as
`measurements' (parameters known to better than $\sim 10$ per cent
accuracy).  Remarkable progress in observations of the temperature
fluctuations in the Cosmic Microwave Background (CMB) has been central
to this scientific transformation (e.g.\ the recent results from the
WMAP satellite, Spergel et al.\ 2003).

The current standard cosmological model (`$\Lambda$CDM') assumes that
the Universe is composed of baryons, cold dark matter (CDM) and
Einstein's cosmological constant $\Lambda$.  The present-day densities
of baryons, (baryons + CDM) and $\Lambda$, as a fraction of the
critical density, are denoted by $\Omega_{\rm b}$, $\Omega_{\rm m}$
and $\Omega_\Lambda$, respectively.  The local rate of cosmic
expansion is described by the Hubble parameter $h = H_0/(100$ km
s$^{-1}$ Mpc$^{-1})$.  Structures in the Universe are assumed to grow
by gravitational amplification of a primordial `power spectrum of
fluctuations' created by inflation, which is taken to be a `scale
free' power law $P_{\rm prim}(k) \propto k^{\, n_{\rm
scalar}}$.\footnotemark \footnotetext{A power law function is `scale
free' because it contains no preferred length scale.  The power law
index $n_{\rm scalar} = 1$ is described as `scale invariant' because
in this case, the matter distribution has the same degree of
inhomogeneity on every resolution scale (is `self-similar').
Inflationary theories of the early Universe predict that this should
be approximately true.}

Within this framework, we begin by reviewing what the CMB tells us
about the cosmological model.  The basic CMB observable is the angular
power spectrum of temperature fluctuations, which consists of a series
of `acoustic peaks'.  We note that the physics of the CMB fluctuations
(i.e.\ the photon-baryon fluid driven by the dark matter potential
wells) depends entirely on the physical densities of matter and of
baryons, $\Omega_{\rm m} h^2$ and $\Omega_{\rm b} h^2$.\footnotemark
\footnotetext{The factor $h^2$ arises because these physical densities
are proportional to the critical density of the Universe, which scales
as $H_0^2$.  The existence today of non-zero curvature or of a
cosmological constant $\Lambda$ is insignificant at recombination.}
However, the {\it projection} of the anisotropy spectrum onto the sky
(i.e.\ the angular locations of the acoustic peaks) also involves the
angular diameter distance to the last-scattering surface.

\begin{itemize}

\item The angular location $\theta_A$ of the first `acoustic peak' of
the temperature power spectrum turns out to be relatively insensitive
to all parameters apart from the {\it curvature of the Universe}.  The
global geometry is {\it close to} spatially flat, although positive
spatial curvatures ($\Omega_k \approx -0.05$) and hence
closed-Universe models remain consistent with all available data
(Efstathiou 2003).  It is important to note that exact spatial
flatness, typically assumed, remains a theoretical prior which demands
observational confirmation.

\item The {\it physical matter density} ($\Omega_{\rm m} h^2 \approx
0.14$) is set by the ratio of the first acoustic peak height to the
large-scale amplitude of the temperature power spectrum.

\item The {\it physical baryon density} ($\Omega_{\rm b} h^2 \approx
0.024$) is fixed by the relative heights of the first and second
acoustic peaks (and is nicely in accord with the value derived from
nucleosynthesis-based arguments).

\item The values of $\Omega_{\rm m} h^2$ and $\Omega_{\rm b} h^2$
together determine the {\it sound horizon} $s$ at recombination, the
characteristic physical scale fixing the spatial position of the first
acoustic peak.  Using the angular location $\theta_A$ of this peak, we
can derive the angular diameter distance $D_A = s/\theta_A$ to the
surface of last scattering (whose redshift $z \approx 1100$ is known
from thermal physics).  {\it Assuming a flat Universe and the
$\Lambda$CDM model}, $D_A(z=1100)$ depends on a combination of
$\Omega_{\rm m}$ and $h$.  Together with the value of $\Omega_{\rm m}
h^2$ known from the first peak height, this breaks the degeneracy
between $\Omega_{\rm m}$ and $h$, yielding $\Omega_{\rm m} \approx
0.3$ and $h \approx 0.7$.

\item The overall normalization of the temperature power spectrum
(i.e.\ the fluctuation strength at $z = 1100$) can be related to the
{\it amplitude of mass fluctuations at redshift zero}, conventionally
defined as the rms mass perturbation in spheres of radius $8 \,
h^{-1}$ Mpc assuming linear evolution\footnotemark \footnotetext{When
the amplitudes of density fluctuations are small (when the overdensity
$\delta \ll 1$), the equations of structure formation may be {\it
linearized} using perturbation theory.  Gravitational growth is said
to be in the {\it linear regime} and may be solved analytically with
relative ease.  Eventually, the fluctuation amplitudes will become too
high and linear theory will break down.} (and denoted $\sigma_8$).  In
a $\Lambda$CDM model, if the matter densities and Hubble constant are
known, then this quantity is still degenerate with the {\it optical
depth to the last-scattering surface} ($\tau$).  The detection of
polarization in the CMB by the WMAP experiment begins to break this
degeneracy by obtaining the value $\tau \approx 0.17$ (hence $\sigma_8
\approx 0.9$).

\item Changing the slope of the primordial power spectrum of mass
fluctuations ($n_{\rm scalar}$) induces a broad-band tilt in the CMB
temperature power spectrum, that may be quantified by e.g.\ the ratio
of the first and third acoustic peaks (note that the quantity
$\Omega_{\rm b} h^2$ only controls the ratio of odd-to-even peak
heights).  Current data favours a roughly scale-invariant primordial
fluctuation spectrum ($n_{\rm scalar} \approx 1$).

\end{itemize}

In summary, according to this crude argument, the CMB yields six
independent quantities (assuming polarization measurements).  For a
flat $\Lambda$CDM model, the WMAP experiment alone measured values of
($\Omega_{\rm m}$, $\Omega_{\rm b}$, $h$, $\sigma_8$, $\tau$, $n_{\rm
scalar}$) to $5-10$ per cent accuracy (see Spergel et al.\ 2003).

Similar values for some of these parameters are yielded by several
independent, non-CMB techniques, for example:

\begin{itemize}

\item Observations of Cepheid variables and secondary distance
indicators to measure $h$ (Freedman et al.\ 2001).

\item Searches for high-redshift supernovae to measure
$\Omega_\Lambda$ (Riess et al.\ 1998; Perlmutter et al.\ 1999).

\item Measurements of the abundance of rich clusters to determine
$\sigma_8$ (e.g.\ Bahcall \& Bode 2003).

\item Measurement of the amplitude of cosmic shear (weak lensing due
to large-scale structure) to constrain $\sigma_8$ (e.g.\ Brown et al.\
2003; Hoekstra, Yee \& Gladders 2002; Jarvis et al.\ 2003; Bacon et
al.\ 2003; Pen et al.\ 2003).

\end{itemize}
The consistency of these results with those obtained from the CMB
demonstrates that -- if the standard cosmological picture is correct
-- these methods are not obviously compromised by systematic errors
{\it at the current level of accuracy}.  The next-generation CMB
experiment, the Planck satellite, is planned to launch in 2007 and
will further refine CMB measurements of these six parameters and
second-order effects via (i) drastically-improved sensitivity on small
scales and (ii) an enhanced polarization capability.\footnotemark
\footnotetext{See {\tt http://www.astro.esa.estec.nl/$\sim$Planck}.
The temperature power spectrum measured by WMAP is already `cosmic
variance limited' (i.e.\ cannot be improved) on large scales up to and
including the first acoustic peak.}

However, if we wish to relax any of the assumptions of our most basic
(flat $\Lambda$CDM) model then, as we will describe, the degeneracies
inherent in the CMB become insuperable and {\it extra information} is
required (Efstathiou \& Bond 1999; Bridle et al.\ 2003).  Just such a
course of action is demanded by today's most pressing cosmological
questions:

\begin{itemize}

\item Is the current rate of cosmic expansion accelerating, as
suggested by recent supernova measurements?

\item If so, does Einstein's cosmological constant (vacuum energy)
exist, despite our inability to understand its value theoretically?
Or, is the accelerating expansion driven by a different form of {\it
dark energy}, such as `quintessence'?  What are the properties of this
dark energy?

\item Can competing models of inflation be discriminated by accurate
measurements of the shape of the primordial power spectrum (e.g.\ by a
`running spectral index' instead of $n_{\rm scalar}$)?

\item Can the tensor (gravitational wave) component be isolated?

\item Is the Universe exactly flat or does it possess a small spatial
curvature $\Omega_k$?

\end{itemize}

In this paper we focus our discussion upon the quest to characterize
the mysterious dark energy, which drives the accelerating cosmic
expansion and contributes the majority of the present-day energy
density of the Universe.  As an explanation of the dark energy,
Einstein's cosmological constant $\Lambda$ is consistent with current
observations but poses {\it tremendous theoretical challenges}.  The
`natural' (quantum-mechanical) value of $\Lambda$ is a factor
$10^{120}$ higher than that measured by astronomers.  This severe
difficulty has motivated the development of alternative models, many
featuring a {\it dynamic} component of dark energy whose properties
evolve with time, such as {\it quintessence}.  These competing models
are essentially untested and require experiments able to measure the
properties of dark energy as a function of cosmic epoch.  The dark
energy model is commonly characterized by its {\it equation-of-state}
$w(z) = P/\rho$ relating its pressure $P$ to its energy density $\rho$
(in units where the speed-of-light $c = 1$).  For Einstein's
cosmological constant, $w(z) = -1$.

The resolution of this problem is the current cosmological frontier,
and has been widely identified as having profound importance for
physics as a whole.  Needless to say, if there were a paradigm shift
in the next decade that alleviated the dark energy problem, then the
SKA experiments described below would be even more important in
confirming and better understanding the new cosmological model.

Can the CMB inform us about dark energy?  Although in the standard
($\Lambda$CDM) model dark energy has a negligible effect at
recombination, the CMB is in fact sensitive to the equation-of-state
of dark energy because the angular positioning of the acoustic peaks
depends on the angular diameter distance to $z=1100$, which is
affected by $w(z)$ (Huey et al.\ 1999).  However, the efficacy of this
probe of $w(z)$ is limited because:

\begin{itemize}

\item Using CMB data alone, there is an immediate degeneracy between
$w(z)$ and $\Omega_{\rm m}$ to produce the same value of
$D_A(z=1100)$.

\item Even given accurate external information about $\Omega_{\rm m}$
and $h$, the quantity $D_A(z=1100)$ can still only constrain a
weighted average of $w(z)$, i.e.\ {\it one degree of freedom}.

\item The precision with which this one degree of freedom can be
measured is further limited because $w(z)$ only influences dynamics at
low redshift, thus its influence upon $D_A(z=1100)$ is small.

\item This lack of sensitivity implies that we can make small
adjustments to the value of the sound horizon $s$ at fixed
$\Omega_{\rm m}$ by changing the mix of baryons and CDM (i.e.\
$\Omega_{\rm b}$) to leave unchanged the angular locations of the CMB
acoustic peaks, which depend on $s/D_A(z=1100)$.

\end{itemize}

\begin{figure*}
\center
\epsfig{file=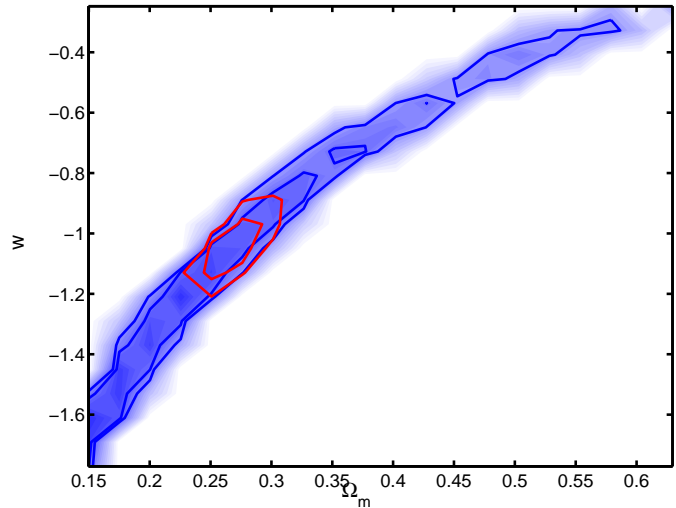,width=9cm,angle=0}
\caption{\small Simulated joint constraints ($68\%,95\%$) on the
matter density $\Omega_{\rm m}$ and a constant dark energy
equation-of-state $w(z) = w_{\rm cons}$, resulting from Planck
observations of the CMB.  We use TT, TE and EE simulated Planck data
and only retain points with $\ell > 50$ to account for uncertainty in
any dark energy perturbations, which may manifest themselves on very
large scales and are strongly dependent on the exact dark energy
model.  The strong degeneracy in the $(\Omega_{\rm m},w_{\rm cons})$
plane arises because the CMB power spectra are approximately invariant
for fixed values of (1) $\Omega_{\rm m} h^2$ and (2) the angular
diameter distance to the last-scattering surface, $D_A(z=1100)$, which
fixes the angular location of the first acoustic peak.  $D_A(z=1100)$
may be held constant by varying $\Omega_{\rm m}$ simultaneously with
$w_{\rm cons}$; the combination $\Omega_{\rm m} h^2$ is kept fixed by
also varying $h$.  The inclusion of an additional tight Gaussian prior
on the Hubble parameter ($h = 0.72 \pm 0.02$) breaks the main
degeneracy, resulting in the tighter inner contours.  However, there
remains a {\it secondary degeneracy} between the value of $w_{\rm
cons}$ and the physical baryon density $\Omega_{\rm b} h^2$.  This
latter quantity can be varied to change the size of the sound horizon
and compensate for changes in $D_A(z=1100)$ due to $w_{\rm cons}$,
keeping the angular peak locations fixed.  The result is that, even if
$\Omega_{\rm m}$ and $h$ are known precisely, Planck data cannot
determine $w_{\rm cons}$ to a precision better than $\pm 0.1$.}
\label{figplanck}
\end{figure*}

This difficulty in measuring a non-Einstein dark energy
equation-of-state $w(z)$ using the CMB alone, even if assumed to be a
constant $w_{\rm cons} \ne -1$, is illustrated by Figure
\ref{figplanck}, in which we simulate the degeneracy between
$\Omega_{\rm m}$ and $w_{\rm cons}$ {\it after Planck data has been
analyzed}.  Even with the addition of a very accurate independent
measurement of $h$, there remains a $\pm 0.1$ scatter in the value of
$w_{\rm cons}$.  We note that changing $w_{\rm cons}$ from $-1$ to
$-1.1$ alters the value of $D_A(z=1100)$ by less than $1\%$.  This is
because the co-moving distance to a given redshift is determined by
the integral
\begin{eqnarray}
x(z) = \int_0^z \frac{dx}{dz'} \, dz' = \int_0^z \frac{c}{H(z')} \,
dz' \nonumber \\ = \int_0^z \frac{(c/H_0) \, dz'}{\sqrt{\Omega_{\rm m}
(1+z')^3 + (1-\Omega_{\rm m})(1+z')^{3(1+w_{\rm cons})}}} \nonumber
\end{eqnarray}
to which dark energy only contributes at very low redshifts ($z
\ltsimeq 1$).  For fixed $\Omega_{\rm m}$ and $h$, the sound horizon
can be changed by a similar fraction by varying $\Omega_{\rm b} h^2$
within its uncertainty range after the Planck experiment.

The prospects for using the CMB to map out any variations in the dark
energy equation-of-state with cosmic time, the key observation
required to discriminate quintessence cosmologies, are even worse.
Even if the exact value of $\Omega_{\rm m}$ were known, the CMB
primary anisotropies only measure a weighted average of $w(z)$ (Saini,
Padmanabhan \& Bridle 2003).  For example, in a simple model $w(z) =
w_0 + w_1 z$, we can always scale $w_1$ with $w_0$ in such a way to
keep $D_A(z=1100)$ fixed and leave the power spectrum of the primary
anisotropies invariant (although there will be {\it second-order}
changes, such as Sunyaev-Zeldovich decrements due to galaxy clusters,
and the late-time Integrated Sachs-Wolfe effect).

We conclude that, due to degeneracies intrinsic to the CMB, the
outstanding questions of the era of precision cosmology {\it demand}
additional experiments of comparable power, with different inherent
parameter degeneracies.  As discussed above, the CMB can only weakly
constrain the dark energy model.  An accurate measurement of $w(z)$
requires {\it extremely precise} cosmological data at lower redshifts,
where dark energy significantly affects the dynamics of the Universe;
for example, per-cent level determinations of the distances to
redshift slices at $z \sim 1$.\footnotemark \footnotetext{Beyond $z
\approx 1.5$, dark energy is expected to have a negligible effect on
dynamics.  This assertion should be tested; but if true, distance
determinations to $z \approx 3$ have limited power for constraining
$w(z)$, and measurements of the Hubble constant at $z \approx 3$ have
no such power.}  What cosmological probes can provide such precision?

\begin{itemize}

\item Observations of distant Type Ia supernovae as `standard candles'
track the luminosity distance as a function of redshift and provide an
`orthogonal constraint' to the CMB.  Supernova surveys will be a
powerful probe of the dark energy model, but we argue that {\it it is
not yet proven} that the experimental systematics can be controlled to
sufficient precision to disentangle the effects of dark energy at the
required accuracy (see Section \ref{secconc}).

\item Redshift surveys of $10^5-10^6$ galaxies in the local Universe
provide an independent measure of $\Omega_{\rm m} h$ via the shape of
the galaxy power spectrum (if $n_{\rm scalar}$, $\Omega_{\rm b}$, the
neutrino masses and any running spectral index are known).  We argue
that the efficacy of this cosmological probe is fundamentally limited
by uncertainties regarding galaxy bias.  In any case, Figure
\ref{figplanck} demonstrates that the CMB cannot produce a very
accurate measurement of dark energy even with the addition of extra
data fixing $\Omega_{\rm m}$ and/or $h$.

\end{itemize}

In our opinion, the critical additional information required to answer
the profound questions posed in this Section will be provided by
experiments measuring:

\begin{itemize}

\item Acoustic oscillatory features in the clustering power spectrum
(Experiment I, Section \ref{secwiggles}).

\item Weak gravitational lensing by large-scale structure, otherwise
known as `cosmic shear' (Experiment II, Section \ref{seclens}).

\end{itemize}

In this paper we argue that radio surveys with the SKA will provide
the ultimate databases for these experiments, assuming that the
telescope is designed with a wide enough field-of-view.

\section{Assumptions of this paper}

\subsection{SKA capabilities}

We adopt in this paper the SKA capabilities given by the current
`Strawman Science Requirements' (Jones 2004) with one key exception:
the instantaneous (1.4~GHz) field-of-view ($FOV$) must exceed the
Strawman value ($FOV \sim 1$ deg$^2$) by more than an order of
magnitude for our proposed experiments.

Abdalla \& Rawlings (2004; hereafter AR04) have considered this
requirement in more detail, concluding that $FOV$ and $\beta$ (the
ratio of the instantaneous bandwidth to the frequency range $0.5-1.4$
GHz required by Experiment I) must obey the relation $(\beta / 0.25)
\times FOV > 40$ deg$^2$ for the survey to have a reasonable ($\sim 1$
yr) duration.  We emphasise that $FOV$ is the field-of-view obtained
at 1.4 GHz (HI at $z=0$) and that the calculations of AR04 assume that
this field-of-view scales with redshift ($\propto \nu^{-2}$ for the
purposes of this study, where $\nu$ is the frequency of
redshifted HI).  AR04 also note the necessity for at least half the collecting
area to reside in a core of diameter $\sim 5$ km for Experiment I, and
that Experiment II requires significant collecting area in the longer
baselines to deliver the sharp angular resolution (at 1.4~GHz)
necessary for measuring cosmic shear.  Both of these requirements are
already in the `Strawman' design.

Using these SKA specifications, and assuming that the field of view
(in units of deg$^{2}$) scales as $\nu^{-2}$, the effective
sensitivity of a `tiled' survey\footnotemark \footnotetext{In which
pointings just overlap at 1.4 GHz, and each receives a 4-hour
integration so that the effective exposure time scales as $(\nu / 1.4~
\rm GHz)^{-2}$.} is sufficient to detect an $M_{\rm HI}^\star \approx
6 \times 10^9 \, \rm M_\odot$ galaxy with a signal-to-noise ratio of 5
out to redshift $z \sim 1.5$ (assuming no evolution in the break of
the HI mass function; see Sec.\ 2 of Rawlings et al., this volume).

In this paper we have not attempted to model those aspects of our
proposed surveys which will be very dependent on the precise SKA
design, such as how the {\it usable} $FOV$ scales with wavelength for
Experiment I, or what dynamic range is achievable for Experiment II;
we note that in practice these limitations might be extremely
significant.  We will also ignore restrictions on `all-sky' surveys
due to finite processing resources which -- even given likely
technological advances over the coming decade -- may still prove
important, given the huge datasets that will be generated.

\subsection{Other assumptions}

For the remainder of this paper we assume, unless stated otherwise, a
fiducial model consisting of a spatially-flat ($\Omega_k = 0$)
Universe with cosmological parameters $h = 0.7$, $\Omega_{\rm m} =
0.3$, $\Omega_{\rm b}/\Omega_{\rm m} = 0.15$, $\sigma_8 = 1$ and
$n_{\rm scalar} = 1$ (broadly consistent with the recent WMAP results,
Spergel et al.\ 2003).  We adopt a fiducial dark energy model
characterized by an equation-of-state $w(z) = w_0 + w_1 z$ where
$(w_0,w_1) = (-0.9,0)$, in order to illustrate the accuracy with which
the SKA will be able to discriminate a general dark energy model from
a cosmological constant model, for which $(w_0,w_1) = (-1,0)$.  Unless
stated otherwise, we neglect contributions from tensors (primordial
gravitational waves) and massive neutrinos.  We assume that structure
in the Universe is seeded by adiabatic Gaussian
fluctuations.\footnotemark \footnotetext{An SKA redshift survey would
provide a useful probe of non-Gaussianity by detecting large numbers
of superclusters, which correspond to high-peak fluctuations in the
primordial density field evolving into the quasi-linear regime.}

\section{Experiment I: HI emission line surveys}
\label{seclss}

The SKA will enable revolutionary progress in the field of large-scale
structure.  In particular, it will map out the cosmic distribution of
neutral hydrogen by detecting the HI 21cm transition at cosmological
distances that are almost entirely inaccessible to current
instrumentation.  Once an HI emission galaxy has been located on the
sky, the observed wavelength of the emission line automatically
provides an accurate redshift, locating the object's position in the
three-dimensional cosmic web.

We simulated the co-moving space densities of HI-emitting galaxies as
a function of redshift by extrapolating the observed local HI mass
function using model `C' of AR04.  We note that, although an integral
property of the HI mass function is constrained by observations of
damped-Ly$\alpha$ systems, considerable uncertainty remains concerning
the redshift-evolution of its break.  However, reasonable models
indicate that the SKA, if designed with a sufficiently wide
field-of-view, can survey the entire visible sky in a year of
operation, locating $\sim 10^9$ HI emission galaxies over a volume
stretching to redshift $z \approx 1.5$ {\it with sufficient number
density that clustering statistics are limited by cosmic variance, not
by shot noise.}  In this case, the precision with which the galaxy
clustering pattern can be measured is limited entirely by the cosmic
volume mapped: the SKA becomes the premier large-scale structure
machine.  In comparison, planned next-generation optical redshift
surveys (e.g.\ the proposed KAOS spectrograph; {\tt
http://www.noao.edu/kaos}) will only cover $\sim 1000$ deg$^2$ at $z
\sim 1$, whereas all-sky photometric redshift surveys (e.g.\
Pan-STARRS; {\tt http://panstarrs.ifa.hawaii.edu}) inevitably result
in a serious degradation of radial power spectrum modes due to the
relatively large redshift error.\footnotemark \footnotetext{Even with
a very optimistic redshift error $\Delta z = 0.02$, a photometric
redshift survey must cover $\approx 10$ times the area of a
spectroscopic experiment to map out the same number of Fourier modes
in the linear regime (Blake \& Glazebrook 2003).}

\subsection{Galaxy power spectrum}

\begin{figure*}
\center
\epsfig{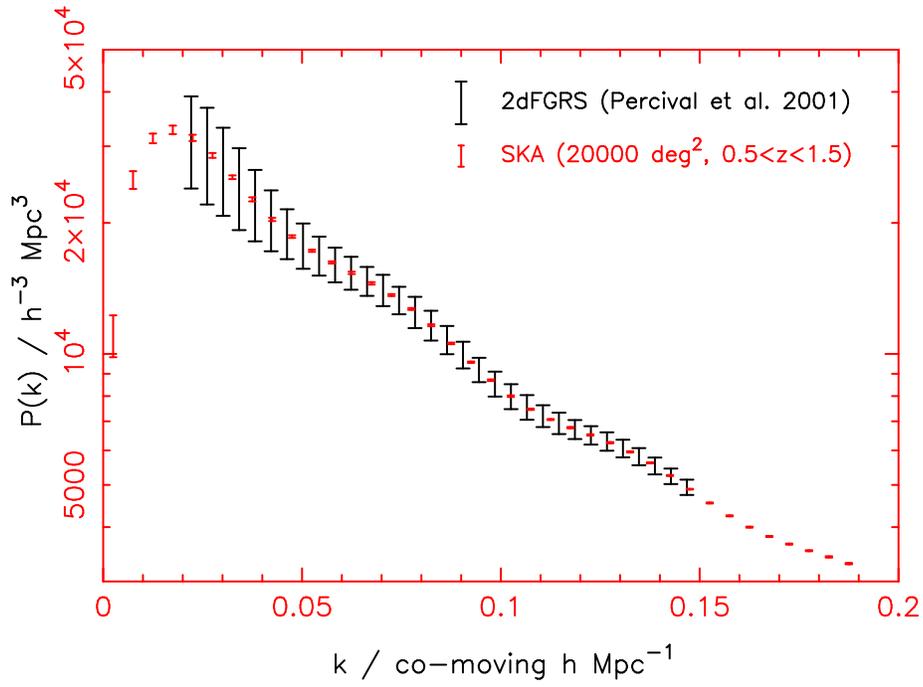}
\caption{\small Simulated measurement of the galaxy clustering power
spectrum by a 1-year SKA survey, compared to the 2dFGRS 100k data
release (Percival et al.\ 2001).  The measurements have been
normalized to the same underlying power spectrum model, keeping the
($1\sigma$) fractional errors in each $k$-bin the same.  For the SKA
data, we assume that the redshift evolution of clustering may be
divided out so that results from different redshift slices can be
combined.  We note that the 2dFGRS $P(k)$ data points are {\it heavily
correlated} in this binning, whereas the SKA survey possesses a much
narrower `$k$-space window function' owing to the vastly greater
cosmic volume probed, implying that the resulting $P(k)$ measurements
are barely correlated.  The SKA clustering data delineates the power
spectrum turnover on large scales ($k < 0.02 \, h$ Mpc$^{-1}$) and the
acoustic oscillations (Section \ref{secwiggles}), as a direct result
of mapping a vastly greater cosmic volume $V$.  The error in a power
spectrum measurement in the cosmic-variance-limited regime scales as
$1/\sqrt{V}$, and $V_{\rm SKA} \approx 500 \, V_{\rm 2dF}$.}
\label{figskapk}
\end{figure*}

The resulting HI redshift survey yields an improvement in power
spectrum precision of between one and two orders of magnitude compared
to the current state-of-the-art (Figure \ref{figskapk}), delineating
features such as the {\it acoustic oscillations} (see Section
\ref{secwiggles}).  In addition, the SKA clustering map traces {\it
clustering modes on large scales beyond the power spectrum turnover}
(i.e.\ on scales $k < 0.02 \, h$ Mpc$^{-1}$) that are entirely
inaccessible to local redshift surveys; together with extra
small-scale modes (to $k \approx 0.2 \, h$ Mpc$^{-1}$) that are hidden
by non-linear clustering at redshift zero.

\begin{figure*}
\center
\epsfig{file=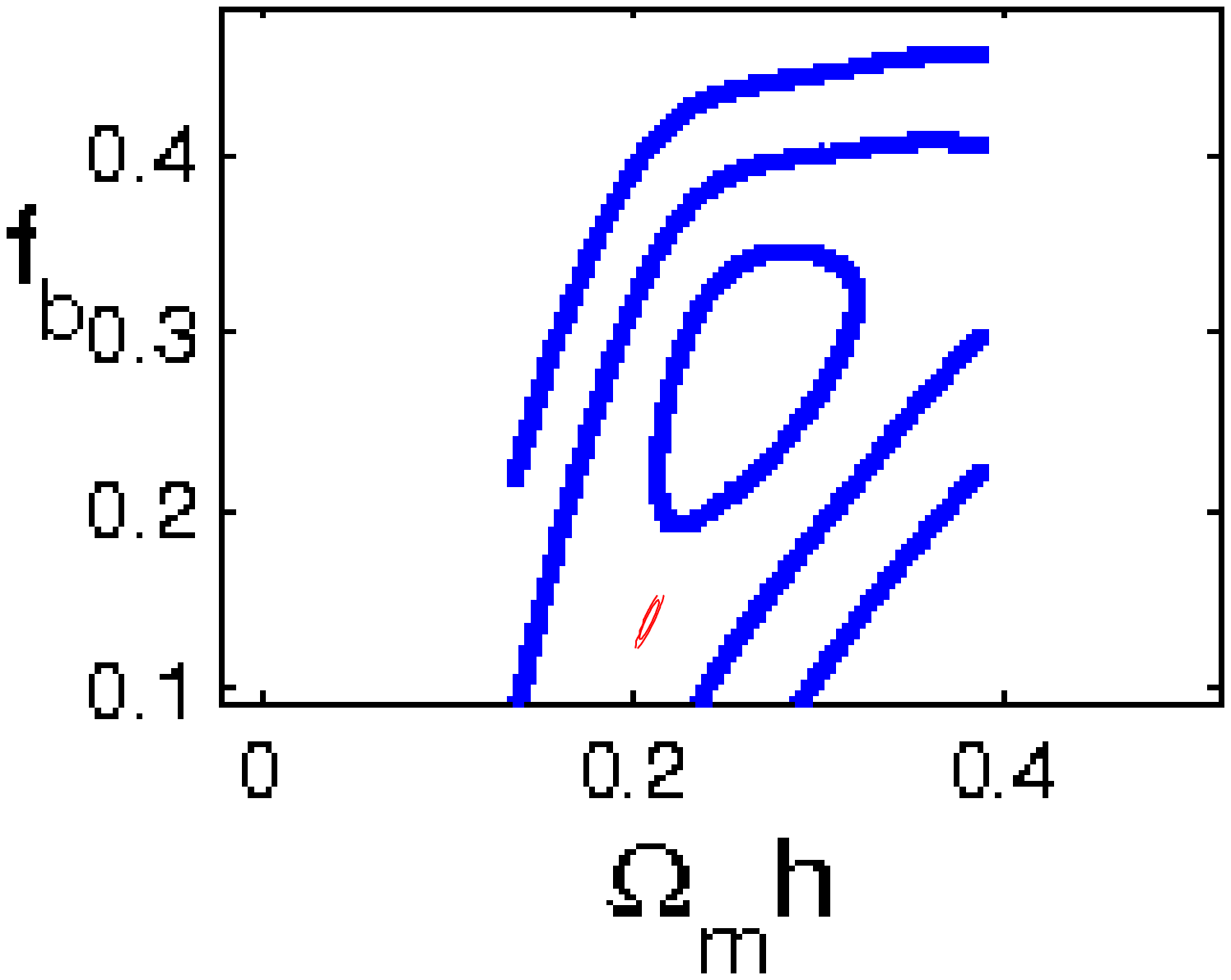,width=12cm,angle=0}
\caption{\small Joint constraints on the cosmological parameters
$(\Omega_{\rm m} h, f_{\rm b} = \Omega_{\rm b}/\Omega_{\rm m})$
resulting from fitting linear theory models (calculated using the CAMB
software; Lewis, Challinor \& Lasenby 2000) to the shape of the SKA
galaxy power spectrum (assuming a linear galaxy biasing scheme).
Contours are shown for the analyzed Sloan Digital Sky Survey (SDSS)
data (Pope et al.\ 2004; $1\sigma,2\sigma,3\sigma$) and for a
simulated SKA HI emission-line survey (inner contour; $2\sigma$).  We
only use SKA clustering data in the linear regime (with $k < 0.2 \, h$
Mpc$^{-1}$).  In order to match the assumptions of the SDSS contours,
we fix $n_{\rm scalar} = 1$ and marginalize over only the Hubble
constant.  Whilst the SKA produces an exquisite delineation of the
power spectrum, we argue that this Figure {\it may be wildly
optimistic} because the bias model cannot be assumed to be linear at
the required level of precision.}
\label{figskaobom}
\end{figure*}

What cosmological information can we glean?  The galaxy clustering
pattern encodes a wealth of data about the underlying constituents of
the Universe and the basic gravitational processes governing the
formation of structure.  In particular, the shape of the power
spectrum depends on a combination of cosmological parameters.  In
Figure \ref{figskaobom} we demonstrate that parameter constraints
resulting from the SKA power spectrum vastly improve upon those
obtained using the Sloan Digital Sky Survey (SDSS).  Note in
particular that, since the acoustic oscillations are detected in the
SKA power spectrum (see Section \ref{secwiggles}), the baryon fraction
is strongly constrained, whereas the SDSS contours are compatible with
zero baryons.

The precise determination of the power spectrum on large scales can be
compared with the predictions of {\it inflationary models}, testing
for effects such as departures of the primordial power spectrum
$P_{\rm prim}(k)$ from a pure power-law.  The narrow window function
in $k$-space (resulting from the vast cosmic volume probed by the SKA)
implies that any {\it sharp features} in $P_{\rm prim}(k)$ are {\it
not smoothed out} and may be detected in a manner not possible using
the CMB.  It has been suggested that the intriguing outliers in the
CMB temperature power spectrum measured by the WMAP satellite (at
$\ell \sim 30$ and $\ell \sim 200$) could be due to unknown physics on
scales smaller than the Planck length, which are imprinted during
inflation (e.g.\ Martin \& Ringeval 2004).  The SKA would delineate
any such effects.  Furthermore, comparison with large-scale CMB modes
can yield information about any {\it tensor contribution} to the CMB
anisotropies, arising from gravity waves in the early
Universe.\footnotemark \footnotetext{Assuming that such a tensor
contribution may be disentangled from any large-scale effects due to
clustering of dark energy, galaxy bias, and the potential presence of
`isocurvature' density fluctuations in the early Universe (in addition
to the usual adiabatic perturbations).}  This is a second potential
means, together with pulsar observations, by which the SKA can be used
as a gravitational wave detector.

However, we wish to argue that the applications of the SKA large-scale
structure measurements discussed above are {\it potentially flawed}
(and at best severely complicated) by our lack of knowledge of the
{\it detailed biasing scheme} linking the observed galaxy distribution
to the underlying pattern of dark matter fluctuations produced by
theoretical models (e.g.\ Coles 2004).  In general, galaxy biasing is
almost certainly stochastic, scale-dependent, non-local and non-linear
(Dekel \& Lahav 1999): the assumption of a simple linear bias,
implicit in Figure \ref{figskaobom}, is likely to be inadequate when
confronted with sub-percent precision power spectrum data.

\subsection{Acoustic oscillations}
\label{secwiggles}

Given the likely complexities of galaxy bias, we seek cosmological
experiments that are less sensitive to the biasing model.  The power
spectrum of galaxies on large scales ($\gtsimeq 30$ Mpc) should
contain a series of small-amplitude {\it acoustic oscillations} of
identical physical origin to those seen in the CMB.\footnotemark
\footnotetext{The acoustic oscillations imprinted in the matter power
spectrum possess significantly lower amplitude than those found in the
CMB power spectrum, because the baryonic component of the matter
distribution, which contains the oscillatory imprint, is sub-dominant
to the CDM component.}  These features result from oscillations of the
photon-baryon fluid before recombination, and encode a characteristic
scale -- the {\it sound horizon at recombination} -- accurately
determined by CMB observations as described in Section \ref{secback}.
This scale can act as a {\it standard ruler} (Eisenstein 2002).  Its
recovery from a galaxy redshift survey depends on the assumed
cosmological parameters, particularly the dark energy model, and thus
constrains that model over a range of redshifts (Blake \& Glazebrook
2003; Seo \& Eisenstein 2003).

\begin{figure*}
\center
\epsfig{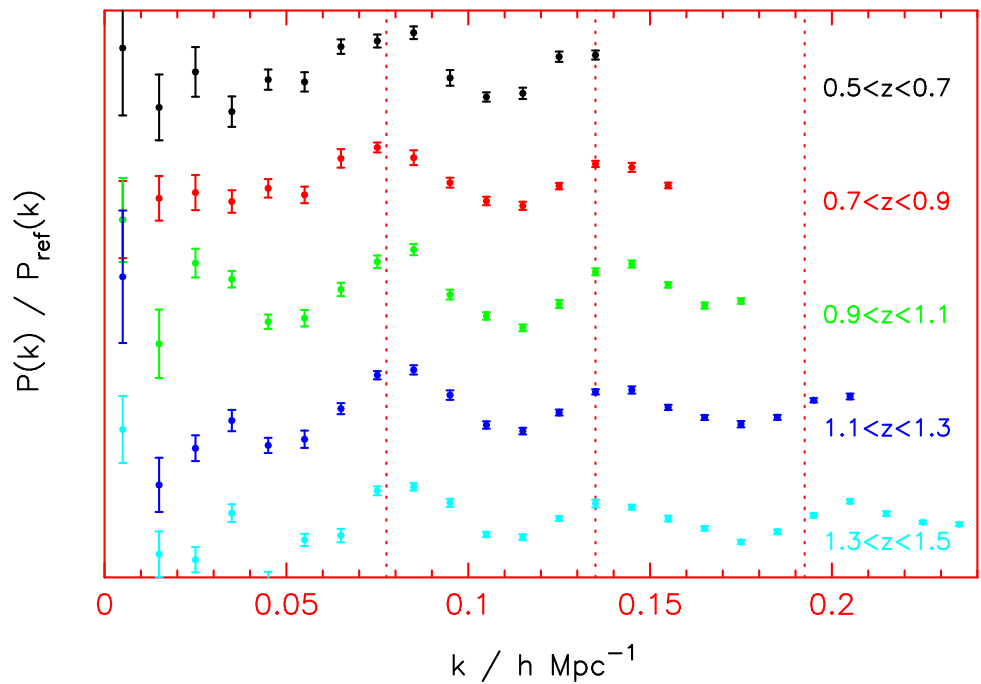}
\caption{\small Simulated galaxy power spectra for an SKA survey of
20,000 deg$^2$, analyzed in redshift slices of width $\Delta z = 0.2$.
Each power spectrum is divided by a smooth polynomial fit, revealing
the sinusoidal imprint of acoustic oscillations, and is shifted along
the $y$-axis for clarity.  As redshift increases, the linear regime
extends to smaller physical scales (larger values of $k$), unveiling
more peaks.  The acoustic oscillation `wavelength' may be used as a
standard ruler.  This is illustrated by the Figure: the true dark
energy model is $w = -1$, but the observer has incorrectly assumed $w
= -0.8$ when constructing the power spectra, and consequently the
recovered acoustic scale disagrees with that observed in the CMB
(represented by the vertical dotted lines).}
\label{figwigpk}
\end{figure*}

The application of this cosmological test does not depend on the
overall shape of the power spectrum, which can be divided out using a
smooth fit, but only on the residual oscillatory signature of the
acoustic peaks.  Hence the method is insensitive to smooth broad-band
tilts in $P(k)$ induced by such effects as a running spectral index,
redshift-space distortions, and complex biasing schemes (e.g.\ based
on haloes; Seljak 2000).  It would be very surprising if the biasing
scheme introduced {\it oscillatory features} in $k$-space liable to
obscure the distinctive acoustic peaks and troughs.  In the absence of
major systematic effects, constraints on the dark energy model are
limited almost entirely by how much cosmic volume one can survey,
rendering this an ideal experiment for the SKA, which can map the HI
galaxy distribution out to $z \approx 1.5$ (where three acoustic peaks
lie in the linear regime).

In Figure \ref{figwigpk} we display simulated power spectra for a
survey covering $20,000$ deg$^2$, divided into redshift slices.  If
the observer applies the wrong dark energy model when converting HI
redshifts to cosmic distances, then the peaks and troughs in the
recovered $P(k)$ lie in the wrong places: the experiment becomes a
powerful probe of dark energy.

\begin{figure*}
\center
\epsfig{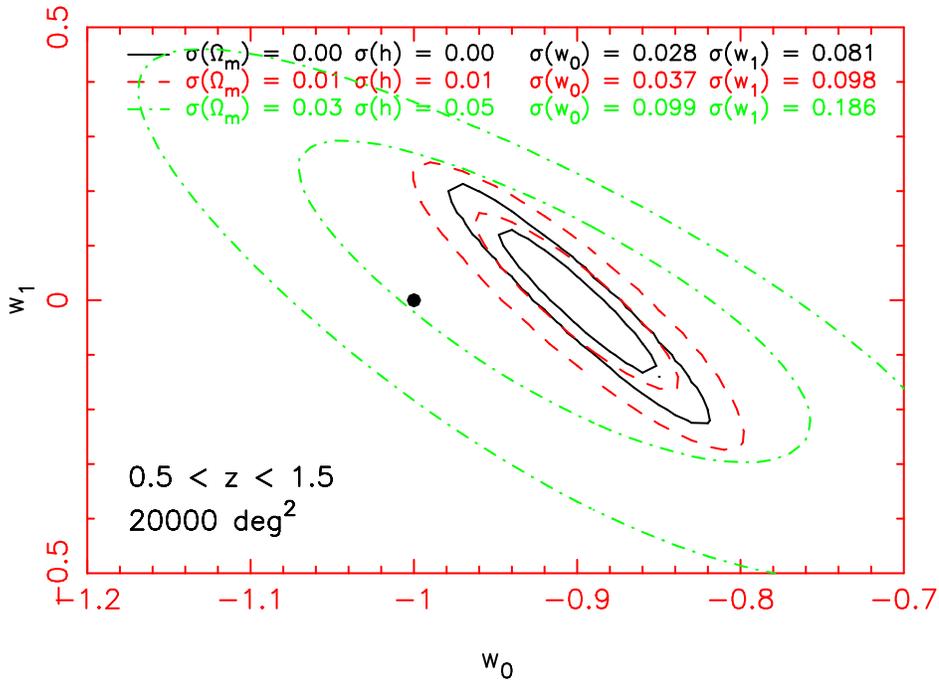}
\caption{\small Constraints on a dark energy model $w(z) = w_0 + w_1
z$ achievable by an SKA survey of the acoustic peaks, illustrated by
sets of $(68\%,95\%)$ contours corresponding to different assumed
priors.  The solid contours correspond to perfect knowledge $\Omega_m
= 0.3$ and $h = 0.7$.  For the dashed contours we impose Gaussian
priors with standard deviations $\sigma(\Omega_m) = 0.01$ and
$\sigma(h) = 0.01$, which are not unrealistic in the SKA epoch.  In
each of these two cases we are able to distinguish a fiducial model
$(w_0,w_1) = (-0.9,0)$ from a cosmological constant (marked by the
solid circle).  Weaker Gaussian priors $\sigma(\Omega_m) = 0.03$ and
$\sigma(h) = 0.05$ (the dot-dashed contours) are inadequate for this
purpose.  Note that there remains a residual degeneracy in the
$(w_0,w_1)$ plane because at a given redshift, approximately the same
cosmology can be obtained by increasing $w_0$ and decreasing $w_1$.
This degeneracy is partially broken by measurements across a range of
redshift slices (and by the independent determinations of $x$ and $x'$
for each redshift).}
\label{figwigw0w1}
\end{figure*}

The standard ruler provided by the acoustic oscillations may be
applied separately to the tangential and radial components of the
power spectrum, resulting in independent measurements of the
co-ordinate distance $x(z)$ to each redshift slice and its rate of
change $x'(z) \equiv dx/dz$, respectively.\footnotemark
\footnotetext{The actual quantities determined by the experiment are
the ratio of $x$ and $x'$ to the sound horizon $s$ at recombination
(which determines the characteristic spatial scale of the acoustic
oscillations).  The value of $s$ is determined by CMB observations,
via $\Omega_{\rm m} h^2$ and $\Omega_{\rm b} h^2$, but is affected to
some extent by uncertainities in $\Omega_{\rm m}$ and $h$ (Figure
\ref{figwigw0w1}).  The tangential component of $P(k)$ measured in a
survey slice at redshift $z$ determines $x(z)$ because this latter
quantity governs the tangential scale of the recovered galaxy
distribution in that redshift slice: displacements $\Delta r$ are
determined by $\Delta r = x \, \Delta \theta$.  The value of $x'(z)$
fixes the radial scale of the recovered distribution: $\Delta r =
(dx/dz) \, \Delta z$.}  We fitted these simulated measurements of
$x(z)$ and $x'(z)$ over a series of redshift slices, assuming a dark
energy model $w(z) = w_0 + w_1 z$.  Figure \ref{figwigw0w1}
illustrates the accuracy of recovery of $(w_0,w_1)$ assuming a
fiducial cosmology $(-0.9,0)$, demonstrating that in this case we can
dismiss a cosmological constant model $(-1,0)$ with high significance.
The accuracy of our prior knowledge of $\Omega_m$ and $h$ affects the
tightness of the dark energy constraints.  These parameters must be
known to standard deviations $\sigma(\Omega_m) = 0.01$ and $\sigma(h)
= 0.01$ if these uncertainties are not to dominate the overall error
in $(w_0,w_1)$.  These demands are not unrealistic on the SKA
timescale: SKA observations of masers in the vicinity of black holes
will determine $h$ to $\sim 1\%$ accuracy (Greenhill, these proceedings)
and CMB data from the Planck satellite will fix $\Omega_m h^2$ to a
similar precision (Balbi et al.\ 2003).

Given that the SKA will be unavailable until $\sim 2020$, what other
projects plan to survey the acoustic peaks?  Two leading contenders
are the multi-object near-infrared/optical spectrographs FMOS and
KAOS, due to commence operation later this decade at the Subaru and
Gemini telescopes, respectively.\footnotemark \footnotetext{KAOS is
still unfunded.}  These projects may perform redshift surveys of a few
hundred deg$^2$, deriving constraints on $(w_0,w_1)$ that are several
times poorer than the $20,000$ deg$^2$ SKA survey presented above (the
error bars roughly scale as $1/\sqrt{f_{\rm sky}}$).  In radio
wavebands, SKA prototypes may be developed such as the HYFAR proposal
(Bunton et al.\ 2003).  The speed of surveying $z = 1$ HI emission
galaxies for such a prototype would lag the SKA by roughly an order of
magnitude; furthermore, the bandwidth may be significantly smaller
than that envisaged for the SKA.

In summary: even if these precursor experiments succeed in detecting a
deviation from Einstein's cosmological constant, the SKA will still be
required for a precise delineation of the new model.  The key
advantage offered by a radio telescope is that it may be designed with
a field of view exceeding an optical spectrograph by a factor of 100:
the SKA can survey {\it all the available volume} by studying the
whole visible sky.  Moreover, SKA sensitivity is necessary and
sufficient for detecting tracers of this volume out to $z = 1.5$.

\section{Experiment II: radio continuum surveys for cosmic shear}
\label{seclens}

Light rays follow geodesics, which bend in the presence of matter.  It
follows that a coherent shape distortion is imprinted in the
distribution of distant background galaxies by mass fluctuations in
the intervening cosmic web: there is a small tendency for galaxies to
be tangentially aligned.  This pattern of {\it cosmic shear} encodes a
vast body of information and, in principle, is a remarkably powerful
cosmological probe.  The shear pattern is directly sensitive to the
mass distribution predicted by the underlying theory, and does not
depend on the complex details of how galaxy light traces mass.
Furthermore, cosmic shear may be used to measure dark energy: which
controls both the {\it angular diameter distances} intrinsic to the
lens equation (e.g.\ the source-lens distance) and the {\it redshift
evolution of the mass fluctuations} producing the lensing.

A successful cosmic shear survey has three leading design
requirements: very high image quality for reliable shape measurements,
a high source surface density ($\gtsimeq 100$ arcmin$^{-2}$) to limit
statistical noise, and a wide survey area to reduce cosmic variance.
Each of these demands is delivered superbly by an SKA radio continuum
survey (Schneider 1999).

The leading systematic for ground-based optical cosmic shear
experiments is the difficulty in quantifying systematic variations in
the point-spread function, due to inevitable changes in the
atmospheric seeing and telescope properties with position and time.
The psf determination is accomplished by observing stars in the field
(of which there are a limited number density).  In comparison, the
point-spread function of a radio telescope is well-determined and
stable (being simply derived from the interferometer baseline
distribution, i.e.\ the synthesized beam).  In addition, the planned
SKA angular resolution ($\approx 0.05$ arcsec at 1.4 GHz) vastly
improves upon that obtainable over wide fields from the ground.

Space-based optical cosmic shear surveys should also exist by the time
the SKA is available.  However, the planned area for these surveys is
$\approx 300$ deg$^2$ (a SNAP-like experiment; Massey et al.\ 2004),
which will be dwarfed by an SKA all-sky continuum survey.  The
fraction of sky $f_{\rm sky}$ mapped is directly related to the
precision of the shear power spectrum measurement via a factor
$1/\sqrt{f_{\rm sky}}$.

An SKA continuum survey will achieve sensitivities of $\sim 30$ nJy in
a 4-hour pointing, far exceeding depths attained by contemporary radio
surveys.  In order to simulate the results of an SKA cosmic shear
experiment, we must assume a model for the radio source populations at
these unprecedented flux densities.  We generated this model from a
combination of two populations: starburst galaxies and quiescent
spiral disks (we ignore for now the Active Galactic Nuclei which
dominate at the highest flux densities but which have too low surface
density to contribute significantly to the weak lensing signal at SKA
sensitivities).  We extrapolated local radio luminosity functions for
these two populations (Sadler et al.\ 2002) as a function of redshift
using a Press-Schechter-based approach, and folded in a prescription
for the physical sizes of each population.  Full details of these
models will be published elsewhere (Abdalla et al., in prep.).

\begin{figure*}
\center
\epsfig{file=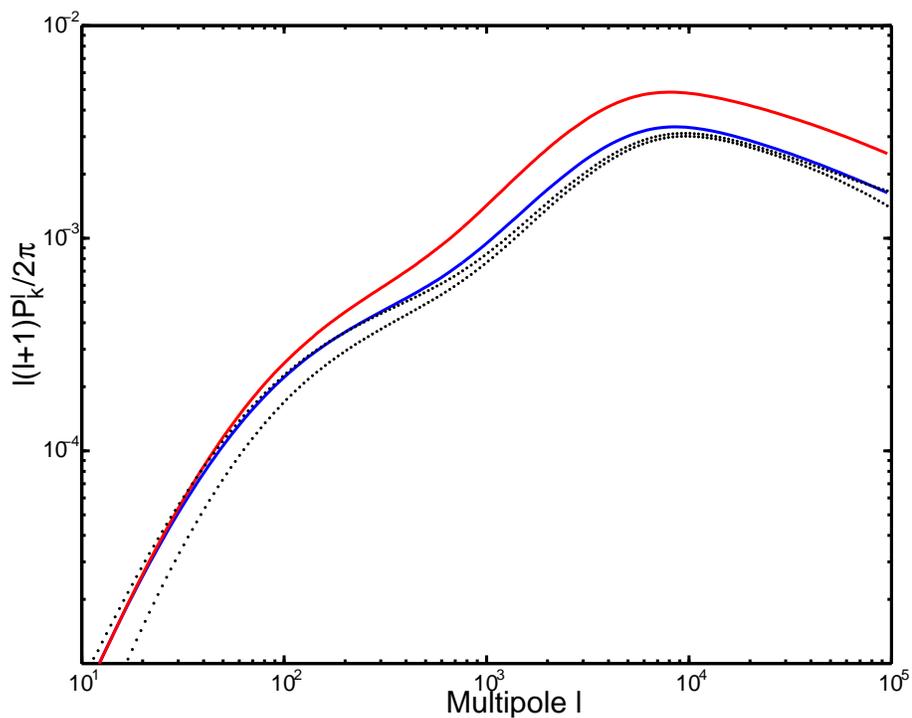,width=12cm,angle=0}
\caption{\small The cosmic shear angular power spectrum measured from
a simulated SKA radio continuum survey. The upper (red) curve is for a
model with $\Omega_{\rm m} = 0.25$ and $w_{\rm cons} = -0.9$; the
lower (blue) curve is for $\Omega_{\rm m} = 0.3$ and $w_{\rm cons} =
-1$; the dotted lines show the $\pm 1\sigma$ range for $\Omega_{\rm m}
= 0.3$ and $w_{\rm cons} = -0.9$.  We assume that the survey contains
a surface density of usable galaxies $n_g = 200$ arcmin$^{-2}$ and
covers one hemisphere ($f_{\rm sky} = 0.5$).}
\label{figlenscl}
\end{figure*}

Using these population models, we simulated radio skies as a function
of SKA integration time and angular resolution.  Following the method
of Massey et al.\ (2004), we then employed standard shape-estimation
software (Kaiser, Squires \& Broadhurst 1995) to recover the shear
dispersion per galaxy arising from intrinsic shapes and from
measurement errors ($\sigma_\gamma$) together with the surface density
of `usable galaxies' ($n_g$) after cuts for objects that are
unresolved or faint.  The values achieved, $\sigma_\gamma \approx 0.2$
and $n_g \sim 500$ arcmin$^{-2}$, are comparable to space-based
optical surveys (Massey et al.\ 2004) whilst covering a survey area
$100$ times larger.  Figure \ref{figlenscl} illustrates the
measurement accuracy of the {\it cosmic shear angular power spectrum}
using these data.\footnotemark \footnotetext{We note that it is also
possible to analyze a radio cosmic shear survey directly in the
$uv$-plane (Chang \& Refregier 2002).}

\begin{figure*}
\center
\epsfig{file=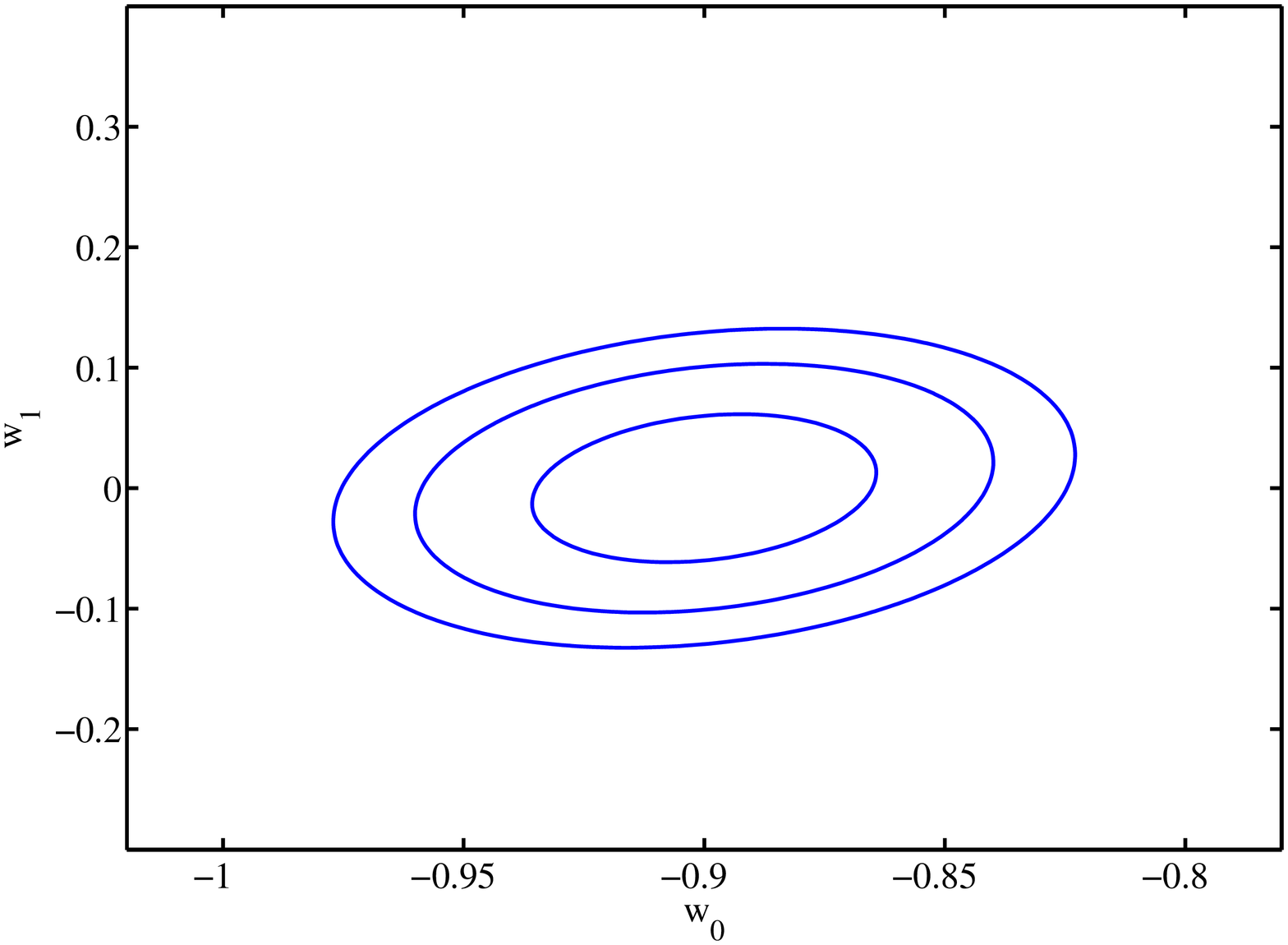,width=9cm,angle=-90}
\caption{ \small $(68\%,95\%,99\%)$ likelihood contours for a dark
energy model $w(z) = w_0 + w_1 z$ from an SKA cosmic shear survey over
a hemisphere, assuming 200 usable galaxies per square arcminute.  We
marginalize over the other cosmological parameters ($\Omega_{\rm m}$,
$\Omega_{\rm b}$, $h$, $n_s$).  These contours are conservative
because no redshift information per galaxy has been included.}
\label{figlensw0w1}
\end{figure*}

The resulting measurements of the dark energy model can be estimated
using the Fisher matrix methodology (e.g.\ Refregier et al.\ 2004).
Figure \ref{figlensw0w1} illustrates joint constraints on $(w_0,w_1)$
using the shear power spectrum up to $\ell = 10^5$.  The analysis
requires models for both the {\it source redshift distribution} (which
can be readily inferred from a deep HI emission-line survey of a
sub-area of the sky) and the {\it linear theory mass power spectrum}
(e.g.\ calculated using CMBFAST; Seljak \& Zaldarriaga 1996) corrected
for non-linear effects in accordance with the prescription of e.g.\
Smith et al.\ (2003).

The power of this experiment may be increased by splitting the lensed
galaxies into two broad redshift bins (Refregier et al.\ 2004).  This
is readily, if roughly, accomplished by separating the radio continuum
sources into two groups determined by detection (or absence) in the HI
emission-line survey of Experiment I.

We note that a cosmic shear survey does not necessarily require
exquisitely-measured galaxy shapes, because there is an irreducible
scatter resulting from the unknown ellipticity of each galaxy {\it
before} shear.\footnotemark \footnotetext{In the ideal cosmic shear
experiment, the unsheared objects would all be circular.}  Instead,
the $0.05$-arcsec SKA resolution is required to (1) resolve starburst
galaxy disks at high redshift and (2) eliminate radio source
confusion.  A potential residual systematic error may lie in {\it
intrinsic alignments} of galaxy disks (owing to structure formation
processes) which may masquerade as coherent shear signal.  For
example, some starburst galaxies may exist in pairs, forming merging
disk systems.  Taking an extremely cautious approach, we assume for
now that starburst galaxies would not be used in the cosmic shear
analysis.  This reduces the surface density by a factor of two, but
the SKA still yields a vast improvement in dark energy constraints
compared to future space-based surveys.

The shear angular power spectrum is the most basic lensing experiment;
we note that the {\it bi-spectrum} (three-point statistic) promises
comparable, statistically-independent constraining power on small
scales (Takada \& Jain 2004).  A third independent approach is to
target {\it the most massive clusters} around which the imprint of
cosmic shear is most significant.  Deep HI follow-up can yield full
background source redshift information, allowing us to cross-correlate
the cosmic shear signal in redshift slices behind the massive lens,
yielding a purely geometric probe of cosmic distances
(`cross-correlation cosmography', e.g.\ Jain \& Taylor 2003).  This
method provides a probe of the angular diameter distance as a function
of redshift independent of the acoustic oscillations in the clustering
power spectrum.  However, a potential systematic error is introduced
by the presence of additional projected masses along the
line-of-sight.

\section{The SKA as the Dark Energy Machine}
\label{secconc}

The current frontier of cosmological research is the characterization
of the properties of the `dark energy' that dominates the energy
density of today's Universe and drives its accelerating expansion.  In
this paper we have focussed on two key experiments -- acoustic
oscillations (Experiment I) and cosmic shear (Experiment II) -- by
which a wide-field SKA can generate measurements of the dark energy
model that are significantly more accurate than any competition by
$\sim 2020$.  Even if precursor surveys detect evidence for a
deviation from Einstein's cosmological constant, the SKA will be
required for a precise delineation of the new model.

\begin{figure*}
\center
\epsfig{file=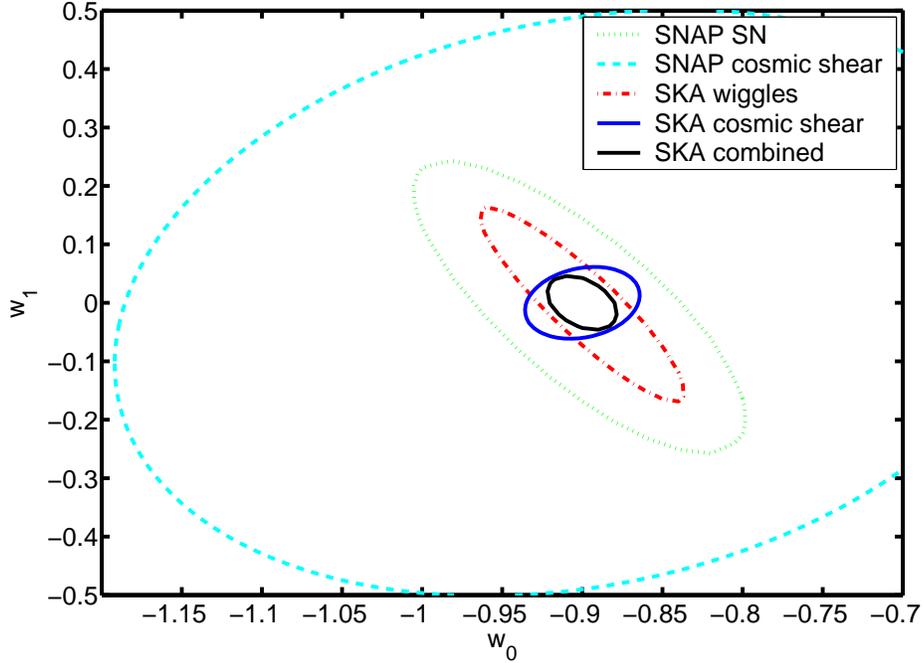,width=9cm,angle=-90}
\caption{\small Comparison of constraints on the dark energy
equation-of-state, parameterized as $w(z) = w_0 + w_1 z$, obtained by
the proposed satellite mission SNAP (using both the supernova and
cosmic shear experiments) and the SKA (via acoustic oscillations and
cosmic shear). The contours represent `$\pm 1 \sigma$' confidence
intervals.  A Gaussian prior upon the matter density,
$\sigma(\Omega_{\rm m}) = 0.01$, was consistently applied to the
supernova and acoustic peaks analyses.  For the latter we also assume
a prior $\sigma(h) = 0.01$ for the Hubble parameter, which could be
obtained by measuring masers using the SKA.  We assumed that the SNAP
supernova component produces measurements of the luminosity distance
to an accuracy of $1.4\%$ in each of 50 redshift bins between $z =
0.1$ and $z = 2$.  The cosmic shear contours for both SNAP and SKA are
conservative because the sample has not been split into two or more
bands in redshift.  This analysis method was shown by Refregier et
al.\ (2003) to improve significantly the constraints from SNAP cosmic
shear, and would have a similar effect on the SKA measurements.}
\label{figallw0w1}
\end{figure*}

In Figure \ref{figallw0w1} we compare overall SKA measurements of a
dark energy model $w(z) = w_0 + w_1 z$ with (arguably) the most
promising competitor: the (currently unfunded) satellite mission SNAP,
planned to launch in $\sim 2014$, whose basic goal is to locate
approximately two thousand supernovae to $z \approx 1.7$ ({\tt
http://snap.lbl.gov}; Aldering et al.\ 2004).\footnotemark
\footnotetext{We note that the planned SNAP mission also includes a
weak lensing component that will yield outstanding space-based image
fidelity.  However, as noted in Section \ref{seclens}, an SKA cosmic
shear survey can cover a area $\sim 100$ times greater with comparable
resolution.}

Observations of distant Type Ia supernovae as `standard candles'
provided the first clear evidence for the accelerating expansion of
the Universe (Riess et al.\ 1998; Perlmutter et al.\ 1999).  Future
high-redshift supernova searches will be a powerful probe of the dark
energy model.  However, we argue that {\it it is not yet proven} that
the dimming effect of cosmic acceleration can be distinguished from
other systematic effects {\it with sufficient precision} to render
this cosmological probe as reliable as Experiments I and II above.
Such systematic effects include changes in the intrinsic properties of
supernovae with galactic environment (e.g.\ metallicity), dust
extinction from both the host galaxy and the Milky Way\footnotemark
\footnotetext{The recent measurement of several supernovae at $z > 1$
by the Hubble Space Telescope (Riess et al.\ 2004) has assuaged this
doubt to some extent by detecting a transition with increasing
redshift from accelerating to decelerating expansion, which would not
occur if the inferred acceleration was an artefact of dust
extinction.}, and the difficulties of photometric calibration
(supernova $K$-corrections, together with the inadequacy of local
`standard stars' such as Vega for providing sub-percent-level flux
calibration).

Many other probes of dark energy have been suggested in the
literature (see e.g.\ Gurvits, this volume).  
We argue that, whilst these probes will provide important
cross-checks of the cosmological model, none can yield the raw
precision or control of systematic errors afforded by our Experiments
I and II.  For example, using SKA (or other) data we can perform {\it
number counts of haloes} as a function of circular velocity and
redshift to infer the cosmological volume element (Newman \& Davis
2000).  But this experiment requires an accurate theoretical
prediction of the number density distribution with circular velocity
(i.e.\ a precise understanding of baryons) together with
precisely-measured rotation curves at high redshift.  Alternatively,
{\it number counts of galaxy clusters} can be utilized (e.g.\ Weller,
Battye \& Kneissl 2002).  The main systematic difficulty here is
connecting the cluster observable (e.g.\ X-ray luminosity or SZ
decrement) to the halo mass (upon which the cluster abundance is
exponentially sensitive).  The frequency of {\it strong gravitational
lensing} (the number of lensing arcs in the vicinity of massive
haloes) is also sensitive to dark energy (Bartelmann et al.\ 2003),
but in manner very dependent on accurate simulations (see also
Koopmans, Browne \& Jackson, this volume)DW.  We argue that
these various experiments will provide insight into the complex
processes of structure formation, rather than into the underlying
cosmological model.

Finally, if matter {\it does not} dominate the dynamics of the
Universe, then gravitational potential wells begin to `decay',
inducing a net shift in the energies of CMB photons passing through
these wells.  This phenomenon is known as the {\it late-time
Integrated Sachs Wolfe (ISW) effect}.  Its influence on the CMB power
spectrum is restricted to large scales ($\ell < 10$) where it is
clouded by cosmic variance.  However, its presence induces a
correlation between CMB anisotropies and the low-redshift matter
distribution, which has been detected using {\it current} radio
surveys (e.g.\ Boughn \& Crittenden 2002) and could be probed further
with the SKA.  Measurement of the late ISW effect is an important
confirmation of the dark energy model, and is sensitive not just to
the evolution of dark energy, but also to its clustering properties
(e.g.\ Weller \& Lewis 2003; Bean \& Dore 2003).

We conclude that, owing to its capacity to undertake cosmic surveys on
an unprecedented scale, the SKA will become the paramount tool for
answering the key questions of the era of precision cosmology.  The
SKA can perform clean and powerful experiments to characterize dark
energy, mapping the acoustic oscillations in the clustering power
spectrum and the patterns of cosmic shear distortion in the radio sky.
Furthermore, SKA data can enlighten other pressing cosmological
questions, pursuing tests of inflationary theories and measuring
spatial flatness with unprecedented precision.

\section*{Acknowledgments}

CB acknowledges Karl Glazebrook for a valuable collaboration
developing simulations of acoustic oscillations.  FA and SB thank Phil
Marshall and Richard Massey for useful discussions about cosmic shear.
CB thanks the SKA Project Office for financial support.  FA thanks
PPARC for a Gemini Research Studentship and the SKA Project Office for
financial support.  SB acknowledges the support of the Royal Society.
SR thanks PPARC for a Senior Research Fellowship and the ATNF for
financial support.

\end{document}